\magnification=1200 \vsize=25truecm \hsize=16truecm \baselineskip=0.6truecm
\parindent=1truecm \nopagenumbers \font\scap=cmcsc10 \hfuzz=0.8truecm
\def\xup{\overline x}
\def\xdo{\underline x}
\def\ydo{\underline y}
\font\tenmsb=msbm10
\font\sevenmsb=msbm7
\font\fivemsb=msbm5
\newfam\msbfam
\textfont\msbfam=\tenmsb
\scriptfont\msbfam=\sevenmsb
\scriptscriptfont\msbfam=\fivemsb
\def\Bbb#1{{\fam\msbfam\relax#1}}
\def\P{$\Bbb P$}

\def\prof{\vrule depth 3.5pt width 0pt}
\def\sba#1#2#3{\prof{\smash{\hbox{\vtop{\hbox{$#1$}%
\hbox{\raise#2pt\hbox{$\mkern#3mu\bar{}$}}}}}}}
\def\sti#1#2#3{\prof{\smash{\hbox{\vtop{\hbox{$#1$}%
\hbox{\raise#2pt\hbox{$\mkern#3mu\tilde{}$}}}}}}}

\def\xsti{\sti x 6 {4}}
\def\asti{\sti a 6 {4}}
\def\Fup{\overline F}
\def\Gup{\overline G}
\def\Fdo{\underline F}
\def\Gdo{\underline G}
\def\uup{\overline u}
\def\vup{\overline v}
\def\udo{\underline u}
\def\vdo{\underline v}
\def\zup{\overline z}
\def\zdo{\underline z}
\def\yup{\overline y}
\def\ydo{\underline y}
\def\Yup{\overline Y}
\def\Ydo{\underline Y}

\null \bigskip
\centerline{\bf A BILINEAR APPROACH TO DISCRETE MIURA TRANSFORMATIONS}

\vskip 2truecm
\bigskip
\centerline{\scap N. Joshi}
\centerline{\sl Dept. of Pure Mathematics}
\centerline{\sl University of Adelaide}
\centerline{\sl Adelaide 5005 Australia}
\bigskip
\centerline{\scap A. Ramani}
\centerline{\sl CPT, Ecole Polytechnique}
\centerline{\sl CNRS, UPR 14}
\centerline{\sl 91128 Palaiseau, France}
\bigskip
\centerline{\scap B. Grammaticos}
\centerline{\sl GMPIB, Universit\'e Paris VII}
\centerline{\sl Tour 24-14, 5${}^e$\'etage}
\centerline{\sl 75251 Paris, France}
\bigskip

\vskip 2truecm \noindent Abstract \medskip
\noindent  We present a systematic approach to the construction of Miura
transformations for
discrete Painlev\'e equations. Our method is based on the bilinear
formalism and we start with the
expression of the nonlinear discrete equation in terms of $\tau$-functions.
Elimination of $\tau$-functions from the resulting system
leads to another nonlinear equation, which is
a ``modified'' version of the original equation.
The procedure therefore yields Miura transformations.
In this letter, we illustrate this approach by
reproducing previously known Miura transformations
and constructing new ones.

\vfill\eject
\footline={\hfill\folio} \pageno=2

The transformations named after Miura relate the solutions of a given
nonlinear equation to
those of some other, also nonlinear equation. The original, most famous,
transformation of this type is
the one relating the solutions of the KdV (Korteweg-deVries) and
modified-KdV equations  [1].
When the
transformation relates the solutions of the {\sl same} equation,
corresponding to different
values of the parameters of the equation, we call them
auto-B\"acklund transformations. Schlesinger
transformations are a particular kind of auto-B\"acklund. In the case of
the Painlev\'e equations
they correspond to special changes of parameters: integer or half-integer
shifts of the
monodromy exponents.

Miura transformations are of particular interest since they reveal
the various interrelations
between nonlinear systems. They are often the first step towards the
construction of the
auto-B\"acklund transformations and Lax pairs. 
In the domain of the (continuous)
Painlev\'e equations, the
systematic derivation and study of Miura, auto-B\"acklund and Schlesinger
transformations has been a subject of
intense activity [2]. It is worth pointing out that the
Schlesinger
transformations of the sixth Painlev\'e equation (P$_{\rm VI}$)
have only recently been
established [3].

As far as the
discrete Painlev\'e equations (d-\P) are concerned, the question of
how to systematically derive interrelationships
is far from settled.
This
is not only due to the fact that the d-\P's have only recently made their
appearance [4] but
also because there are so many of them. In a series of works we have
obtained the Miura
transformations for several d-\P's [5,6,7,8]. The approach we used was
based on the observation
that all the Miura transformations (in perfect analogy to the continuous
case) are rational
expressions involving the ratio of ``Riccati forms'':
$$y={\alpha\xup x+\beta\xup+\gamma x+\delta\over \epsilon\xup x+\zeta\xup+\eta
x+\theta}\eqno(1)$$ where $x=x(n)$, $\xup=x(n+1)$, $\xdo=x(n-1)$ and
similarly for $y$.
Postulating such a form (1),
one starts with $x$ satisfying a given d-{\P}  and
obtains another
d-{\P} for $y$, by determining the coefficients $\alpha,\beta,\dots\theta$.
The procedure is quite
cumbersome, in particular if one works with the full generality
of (1). In
practice, a good
deal of intuition is needed in the derivation of the
Miura transformations of various
d-\P's.

In this letter we shall present a new, systematic, approach to the
construction of Miura transformations of
discrete Painlev\'e equations. It is based on the bilinear formalism. As we
have shown in our
previous works, the $\tau$-functions entering the bilinear
transfomations are defined by
the singularity structure of the discrete equation. In particular,
they are determined by
the singularity pattern contained in an orbit that becomes nearly
singular
[9]. Moreover, the number of $\tau$-functions corresponds to the number
of types of singularity that can occur.
The $\tau$-functions are known to play a major role in the Miura,
auto-B\"acklund and Schlesinger
transformations of the d-\P's  and, in particular, in their self-dual
description [10].
For the derivation of the Miura transformations, as we shall see below,
we will need only their discrete logarithmic derivatives.

Let us start with an example based on a well-known (and well-studied)
discrete Painlev\'e
equation, the ``standard'' d-P$_{\rm II}$:
$$\xup+\xdo={zx+a\over 1-x^2},\eqno(2)$$ where $z=\alpha n+\beta$. The
bilinearisation of (2)
was given in [9,11]. We have argued there that, due to the singularity
structure of (2) i.e.
$\{\pm1,\infty,\mp1\}$, two $\tau$-functions, $F$ and $G$ are needed and we
proposed the
transformation:
$$x=1-{\Fup\Gdo\over FG}=-1+{\Fdo\Gup\over FG}.\eqno(3)$$ As we hinted above
and since our aim
is not the bilinearisation of d-P$_{\rm II}$, we can simplify our
calculation by introducing
the auxiliary variables $u=F/\Fup$, $v=\Gdo/G$. Thus we have
$$x=1-{v\over u}=-1+{\,\udo\,\over \vup}.\eqno(4)$$ The important ``trick''
in the derivation of
the ``modified'' equation is to eliminate one of the two
$\tau$-functions or, equivalently, one of the $u,v$ appearing in the
expression (4), for
instance
$u$. (Note that one has to up- and/or down-shift both (2) and (4) in order
to have enough
equations for this elimination.) The net result is a {\sl homogeneous}
5-point equation for
$v$. Given the homogeneity one can integrate it by putting $v/\vup=w$. The
resulting equation
involves only products of $w\overline w$ (up- or down-shifted). We can
integrate the equation
once more by putting $w\overline w=\omega$. We then obtain for $\omega$ the
three-point
equation:
$$(\overline\omega+\omega-\overline z)( \omega+\underline\omega-z)
={(2\omega-a-z)(2\omega+a-\overline z)\over\omega} .\eqno(5)$$ 
This is a discretization [12]
of equation XXXIV of the Painlev\'e-Gambier list [13]
which is well known to be related to P$_{\rm II}$ by a Miura transformation.
The standard form
of d-P$_{\rm 34}$ can be recovered by putting $\omega=y+z/2+\alpha/4$, 
where
$\alpha=z-\zdo$. We then find:
$$(\overline y+y)( y+\underline y)={4y^2-(a-\alpha/2)^2\over
y+z/2+\alpha/4}.\eqno(6)$$					
This equation can be regarded as a modified version of dP$_{\rm II}$
if we can explicitly construct the Miura transformation relating the two.
We do this by
retracing the above sequence of calculations.
{}From the definitions of $\omega$, $w$, and $v$ we have:
$$y={\,v\,\over \overline{\vup}}-{z \over 2}-{\alpha\over 4}\eqno(7)$$ 
Using (4) we
find
$v=(1-x)u$ and $\vup=\udo/(1+\xup)$. This leads to
$$y=(1+{\xup})(1-x)-{z \over 2}-{\alpha\over 4},\eqno(8)$$ 	
i.e. the
well-known Miura
transformation of d-P$_{\rm II}$ (see eq (9a) of [5]).
Using (8) and the d-P$_{\rm II}$ equation
(2) we can
construct in a straightforward way the inverse of the Miura
transformation.

Another most interesting equation is the alternate d-P$_{\rm II}$, studied
in detail in [14]:
$${z\over 1+x\overline x}+{\zdo\over 1+x\underline x}={1\over x}-x+\zdo+\mu
\eqno(9)$$ This
equation has two singularity patterns $\{0,\infty\}$ and $\{\infty,0\}$ which
suggest the
following expression in terms of $\tau$-functions:
$$x={F\Gdo\over \Fdo G}.\eqno(10)$$ As in the previous case, we work with
the discrete
logarithmic derivatives of the $\tau$-functions:
$u=\Fdo/F$, $v=\Gdo/G$. In [14] we have shown that for the bilinearization
 of (9) we must introduce a separation through a first bilinear condition:
$$\Fup\Gdo+\Fdo\Gup=zFG.\eqno(11)$$ This equation will also be our starting
point here.
Rewriting (11) in terms of $u$, $v$, we have:
$${v\over \overline u}+{u\over \vup}-z=0 \eqno(12)$$ Using (12) and (9) we
can eliminate $u$.
The result is a 4-point homogeneous equation in $v$ which we can integrate
once through
$v/\vup=y$. We obtain thus
$$(\yup-\ydo)^2y^4+2z^2(\yup+\ydo)y^2(1-y)
+z^2\Bigl(4y^3+y^2(z^2-\mu^2-4)-2yz^2+z^2\Bigr)=0\eqno(13)$$  
which is
precisely the equation
obtained in [14]. We remark that (13) is a three-point correspondence
and
not a simple mapping.
As we have remarked in [14], this indicates that the actual form of the
modified equation is a
4-point one and equation (13) is its integral.
Again, the Miura transformation
$$y = {zx\over 1+x \xup} $$
follows from the above calculations.

Up to now we have used our method on equations for which the result was
already known, having
been obtained by a different approach. It is interesting to see whether we
can use it to obtain
new results corresponding to equations that have not yet been studied.
Two examples will be
presented in what follows.

We start with a $q$-P$_{\rm II}$ equation which was first identified in [15]:
$$(\overline x x-1)( x\underline x-1)={z\zdo\over a^2}{x\over
x-az}\eqno(14)$$ where
$z=z_0\lambda^n$ (and $\zdo=z/\lambda$, $\zup=z\lambda$). The singularity
structure of (14) is
easily obtained. Two patterns exist, namely $\{az,\infty,0\}$ and
$\{0,\infty,az\}$, a fact
that suggests the introduction of two $\tau$-functions $F$ and $G$. We are
thus led to the
ansatz:
$$x={\Fup\Gdo\over FG}=az+{\Fdo\Gup\over FG}\eqno(15)$$ Using the
$\tau$-functions we can
rewrite (14) in the form:
$$(\overline {\Fup} \Gdo -F\Gup)({\Fup}\underline{ \Gdo }-\Fdo
G)={z\zdo\over a^2} \Fup G F
\Gdo\eqno(16)$$ which suggests immediately the separation:
$$\overline {\Fup} \Gdo -F\Gup={z\over a}  \Fup G.\eqno(17)$$ As in the
previous case, we
introduce $u=\Fup/F$ and $v=G/\Gdo$ and rewrite equations (15) and (17) as:
$${u\over v}=az+{\vup\over \udo}\eqno(18a)$$
$${\uup\over v}={z\over a} +{\vup\over u}\eqno(18b)$$ Next, we eliminate
$v$ and introduce
$y=\uup/u$. We obtain for the latter the equation:
$$(y\yup-1)(y\ydo-1)={z\zdo\over y}(y-a^2)(y-{1\over a^2})\eqno (19)$$
Equation (19) is indeed
a $q$-form of equation P$_{\rm 34}$ as was shown in [15].

Our final example will be again based on a $q$-P$_{\rm II}$ equation:
$$\xup\xdo=az{x+z\over x(x-1)}\eqno(20)$$ which was identified in [15]. The
singularity
patterns of (20) are $\{1,\infty,0\}$ and
$\{0,\infty,1\}$ suggesting the following ansatz in terms of two
$\tau$-functions $F$, $G$:
$$x={\Fup\Gdo\over FG}=1+{\Fdo\Gup\over FG}.\eqno(21)$$ As previously, we
introduce $v=\Gdo/G$
and $u=F/\Fup$. Expressing (20) and (21) in terms of $u$,
$v$, we eliminate  $v$. Next we introduce the nonlinear variable:
$$y={\,\udo\,\over \uup}\eqno(22)$$ and obtain the equation
$$(y\ydo-az^2)(y\yup-a\lambda^2z^2)=-\lambda za(y-az)(y-\lambda
z).\eqno(23)$$ Equation (23)
can be  brought into canonical form if we put $y=Yz\sqrt{\lambda a}$. We
thus obtain
$$(Y\Ydo-1)(Y\Yup-1)=-{1\over z}(Y-c)(Y-{1\over c}).\eqno(24)$$ where
$c=\sqrt{a/ \lambda }$.
The explicit construction of the Miura transformation is straightforward.
{}From (22), we find:
$$y=\xup(x-1)\eqno(25a)$$ and using (20) or (24) we obtain the second half
of the Miura:
$$x={y\ydo-az^2\over az-y}.\eqno(25b)$$ It is interesting here to actually
use these
transformations in order to construct the Schlesinger transformation of the
$q$-P$_{\rm II}$
(20). We start by introducing another Miura, through the elimination of $u$
rather than $v$. We
find:
$$\psi={\,\vdo\,\over \vup}=\xdo(x-1)\eqno(26)$$ and the canonical equation
is obtained for
$\Psi$ where $\psi=\Psi z\sqrt{a/\lambda}$:
$$(\Psi\overline\Psi-1)(\Psi\underline \Psi-1)=-{1\over
z}(\Psi-\gamma)(\Psi-{1\over
\gamma})\eqno(27)$$ where $\gamma=\sqrt{\lambda a}=\lambda c$. Here is how
one constructs the
auto-B\"acklund: we start from $x$ associated to a given $a$. We construct
$\psi$ (and $\Psi$)
related to
$\gamma=\sqrt{\lambda a}$. Then since $Y$ and $\Psi$ obey the same equation
for different
parameter we can consider that $\Psi$ is indeed a $Y$ provided we change
the parameter of (24)
to $\tilde c=\gamma=\sqrt{\tilde a/ \lambda }$ which means that $\tilde
a=\lambda^2 a$.
 From $Y=\Psi$ we can easily compute the relation between $y$ and $\psi$:
$y=\lambda^2 \psi$
(because $Y$ is computed at $\tilde a=\lambda^2 a$). Combining now (26) and
(25b) we find:
$$\tilde x={az(\lambda \xdo(x-1)-z)\over x(az-\xdo(x-1))}={z(\lambda
a(x+z)-x\xup)\over
x(x(\xup-1)-z)}\eqno(28)$$ the two expressions in the r.h.s. of (28) being
perfectly equivalent
because $x$ satisfies (20). In an analogous way we can obtain the Schlesinger:
$$\xsti={az( \xup(x-1)/\lambda-z)\over x(az-\xup(x-1))}={z(
a(x+z)/\lambda-x\xdo)\over
x(x(\xdo-1)-z)}\eqno(29)$$ where $\xsti$ is  a solution of (20) for
$\asti=a/\lambda^2$.

Let us summarize our findings here. In this letter we have adressed the
problem of the
systematic construction of the Miura transformations for discrete
Painlev\'e equations and
obtained a new approach. Our method is based on the bilinear formalism and
the construction of
the modified equation proceeds through the elimination of one of the
$\tau$-functions. The
method can be applied to both difference- and $q$- discrete equations. Its
usefulness lies in
the fact that using the  Miura transformations we can proceed to the
construction of the
auto-B\"acklund and Schlesingers in a straightforward way. In a future
publication we expect to
apply this approach to the long list of d- and $q$-\P's that have not yet been
examined.

\noindent \bigskip {\scap Acknowledgement}\smallskip
The research reported in this letter was partially supported by the
Australian Research Council.

\noindent \bigskip {\scap References}
\smallskip
\item{[1]} R. M. Miura, J. Math. Phys. 9 (1968) 1202.
\item{[2]} A.S. Fokas and M.J. Ablowitz, J. Math. Phys. 23 (1982) 2033,
\item{} A.S. Fokas, U. Mugan and M.J. Ablowitz, Physica D 30 (1988) 247,
\item{} A.S. Fokas,  and X. Zhou, Comm. Math. Phys. 144 (1992) 601.
\item{} A.P. Bassom, P.A. Clarkson, and A.C. Hicks,
	Stud. Appl. Math. 95 (1995) 1.
\item{} A.E. Milne, P.A. Clarkson, and A.P. Bassom,
	Stud. Appl. Math. 98 (1997) 139.
\item{[3]} U. Mugan and A. Sakka, J. Math. Phys. 36 (1995) 1284.
\item{[4]} A. Ramani, B. Grammaticos, J. Hietarinta,	Phys. Rev. Lett. 67
(1991) 1829.
\item{[5]} A. Ramani, B. Grammaticos, Jour. Phys. A 25 (1992) L633
\item{[6]} M. Jimbo, H. Sakai, A. Ramani, B. Grammaticos, Phys. Lett. A
217 (1996) 111
\item{[7]} 	K. M. Tamizhmani, B. Grammaticos, A. Ramani,
	Lett. Math. Phys. 29 (1993) 49
\item{[8]} K.M. Tamizhmani, A. Ramani, B. Grammaticos, Y. Ohta, Lett. Math.
Phys. 38 (1996) 289.
\item{[9]} 	A. Ramani, B. Grammaticos, J. Satsuma, Jour. Phys. A 28
(1995) 4655.
\item{[10]} A. Ramani, Y. Ohta, J. Satsuma and B. Grammaticos, {\sl
Self-duality and Schlesinger
chains for the asymmetric d-P$_{\rm II}$ and $q$-P$_{\rm III}$ equations},
to appear in Comm.
Math. Phys.
\item{[11]} J. Satsuma, K. Kajiwara,  B. Grammaticos, J. Hietarinta, A.
Ramani, Jour. Phys. A 28
(1995) 3541.
\item{[12]} A. S. Fokas, B. Grammaticos, A. Ramani, J. of Math. Anal. and
Appl. 180 (1993) 342.
\item{[13]} E.L. Ince, {\sl Ordinary differential equations}, Dover, New
York, 1956.
\item{[14]} F. Nijhoff, J. Satsuma, K. Kajiwara, B. Grammaticos, A. Ramani,
Inv. Probl. 12
(1996) 697.
\item{[15]} A. Ramani, B. Grammaticos, Physica A 228 (1996) 160.
\end